\documentstyle[aps,prd, preprint]{revtex}

\begin{document}
\title{On a generalized gravitational Aharonov-Bohm effect}
\author{Geusa de A. Marques$^{1,2}${\thanks{%
gmarques@fisica.ufpb.br}} and Valdir B. Bezerra$^{1}${\thanks{%
valdir@fisica.ufpb.br}}.}
\maketitle

\begin{center}
$^{1.}${\it {Departamento de F\'{\i}sica, Universidade Federal da
Para\'{\i}ba,}}

{\it {Caixa Postal 5008,  Jo\~{a}o Pessoa, Pb, Brazil.}}

$^{2.}${\it {Departamento de F\'{\i}sica, Universidade Estadual da
Para\'{\i}ba,}}

{\it {Av. Juv\^encio Arruda S/N, Campina Grande, Pb, Brazil.}}
\end{center}

\abstract

A massless spinor particle is considered in the background gravitational
field due to a rotating massive body. In the weak field approximation it is
shown that the solution of the Weyl equations depend on the angular momentum
of the rotating body, which does not affect the curvature in this
approximation. This result may be looked upon as a generalization of the
gravitational Aharonov-Bohm effect.\

PACS numbers: 04.20.-q, 0.4.62.+v, 0.4.90.+e

\vskip 3.0 cm

In classical electromagnetism, a charged particle is influenced only by the
electromagnetic fields at the location of the particle. There is no physical
influence upon a charged particle when it passes through a field-free region.

At the quantum level, however, the behavior of a charged particle is
modified by the action of electromagnetic field confined to a region from
which the particle is excluded. This nonlocal phenomenon in which electrons
are physically influenced by electromagnetic fields not experienced by them
is the well-known electromagnetic Aharonov-Bohm effect\cite{y}. This can be
understood as a manifestation of a locally trivial vector potential of a
gauge field which can leads to observable effects in a nontrivial topology
through the phase of the wave function of a charged particle. The
electromagnetic Aharonov-Bohm effect\cite{y} has been widely studied
theoretically\cite{m}, and confirmed experimentally in recent years\cite{a}.

In a metric theory of gravitation, a gravitational field is frequently
related to a nonvanishing Riemann curvature tensor. However, the presence of
localized curvature can have effects on geodesic motion and parallel
transport in regions where the curvature vanishes. The best known example of
this nonlocal effect is provided when a particle is transported around an
idealized cosmic string along a closed curve in which case the string is
noticed at all. This situation corresponds to the gravitational analogue\cite
{4} of the electromagnetic Aharonov-Bohm effect\cite{y}. These effects are
of topological origin rather than local. The electromagnetic Aharonov-Bohm
effect represents a global anholonomy associated with the electromagnetic
gauge potentials. Its gravitational counterpart may be viewed as a
manifestation of nontrivial topology of spacetime. It is worth to call
attention to the fact that differently from the electromagnetic
Aharonov-Bohm effect which is essentially a quantum effect, the
gravitational analogue appears also at a \  purely classical context. Thus,
in summary, the gravitational analogue of the electromagnetic Aharonov-Bohm
effect is the following: particles constrained to move in a region where the
Riemann curvature tensor vanishes may exhibit a gravitational effect arising
from a region of nonzero curvature from which they are excluded. This effect
may be viewed as a manifestation of the nontrivial topology of spacetime. In
a more general sense, particles constrained to move in a region where the
Riemann curvature is nonzero, but does not depend on certain parameters such
as velocity,  like in the case of moving mass currents\cite{5},  or the
angular momentum of a rotating body\cite{6}, in both examples in the weak
field approximation, may exhibit gravitational effects associated with each
one of these parameters in the respective cases. This kind of gravitational
effect we are calling generalized gravitational Aharonov-Bohm effect.

The existence of a gravitational analogue of the electromagnetic
Aharonov-Bohm effect was first pointed out by in the end of the sixties of
the last century[7-12]. Studies concerning this subject has also received
attention by many authors from that time up to now\cite{13}.

Effects analogous to the electromagnetic Aharonov-Bohm effect\ exist in the
classical context like the Sagnac effect in general relativity$\cite{14}$
which consists of a phase shift between two beams of light traversing in
opposite directions the same path around a rotating mass distribution. With
regard to the quantum theory we can mention the experiments using a neutron
interferometer to measure \ the Newtonian gravitational effects on the phase
difference of two neutron beams[15,16]$.$

In this paper we consider a massless spinor particle in the spacetime of a
rotating massive body and study the influence of the angular momentum of a
rotating source on the behaviour of a particle. Gravitational effects are
taken into account in the weak field approximation in which case the
curvature does not depend one the angular momentum of the source. The
generalized gravitational analogue of the electromagnetic Aharonov-Bohm
effect set up is this scenario. It is show that in this case there is a
nonlocal effect of the angular momentum which is coded in the solution of
the Weyl equations.

To begin with let us write down the covariant Dirac equation in a curved
spacetime, for a massless spinor field $\Psi $, which is given by

\begin{equation}
\left[ i\gamma ^{\mu }\left( x\right) \partial _{\mu }+i\gamma ^{\mu }\left(
x\right) \Gamma _{\mu }\left( x\right) \right] \Psi \left( x\right) =0,
\label{1}
\end{equation}
where $\gamma ^{\mu }\left( x\right) $ are the generalized Dirac matrices
and are given in terms of the standard flat space Dirac matrices $\gamma
^{\left( a\right) }$ as
\begin{equation}
\gamma ^{\mu }\left( x\right) =e_{\left( a\right) }^{\mu }\left( x\right)
\gamma ^{\left( a\right) }  \label{2}
\end{equation}
where $e_{\left( a\right) }^{\mu }\left( x\right) $ are tetrad components
defined by
\begin{equation}
e_{\left( a\right) }^{\mu }e_{\left( b\right) }^{\nu }\eta ^{\left( a\right)
\left( b\right) }=g^{\mu \nu }.  \label{3}
\end{equation}
Here $\mu $, $\nu $ are curved global spacetime indices and $a$, $b$ are
flat tetradic spacetime indices.

The product \ $\gamma ^{\mu }\Gamma _{\mu }$ that appears in Eq. (\ref{1})
can be written as \cite{17}

\begin{equation}
\gamma ^{\mu }\left( x\right) \Gamma _{\mu }\left( x\right) =\gamma ^{\left(
a\right) }\left( A_{\left( a\right) }\left( x\right) +i\gamma ^{\left(
5\right) }B\left( x\right) \right) ,  \label{4}
\end{equation}
with $\gamma ^{\left( 5\right) }=i\gamma ^{\left( 0\right) }\gamma ^{\left(
1\right) }\gamma ^{\left( 2\right) }\gamma ^{\left( 3\right) }$ and $%
A_{\left( a\right) }$ and $B_{\left( a\right) }$ given by
\begin{equation}
A_{\left( a\right) }=\frac{1}{2}\left( \partial _{\mu }e_{\left( a\right)
}^{\mu }+e_{\left( a\right) }^{\rho }\Gamma _{\rho \mu }^{\mu }\right)
\label{5}
\end{equation}
and
\begin{equation}
B_{\left( a\right) }=\frac{1}{2}\epsilon _{\left( a\right) \left( b\right)
\left( c\right) \left( d\right) }e^{\left( b\right) \mu }e^{\left( c\right)
\nu }\partial _{\mu }e_{\nu }^{\left( d\right) },  \label{6}
\end{equation}
where $\epsilon _{\left( a\right) \left( b\right) \left( c\right) \left(
d\right) }$ is the completely antisymmetric fourth-order unit tensor.

Now, let us consider the spacetime generated by an infinitely long,
infinitely thin massive cylindrical shell rotating around its axis. In the
weak field approximation the metric reads$\cite{6}$%
\begin{equation}
ds^{2}=-\left( 1-\frac{a(\rho )}{2}\right) dt^{2}+\left( 1+\frac{a\left(
\rho \right) }{2}\right) \left( d\rho ^{2}+\rho ^{2}d\varphi
^{2}+dz^{2}\right) +2bdtd\varphi ,  \label{7}
\end{equation}
where
\begin{equation}
a\left( \rho \right) =-8\mu \Theta \left( \rho -\rho _{0}\right) \ln \left(
\frac{\rho }{\rho _{0}}\right)  \label{8}
\end{equation}
and
\begin{equation}
b\left( \rho \right) =4\mu w\rho _{0}\left[ \frac{\rho ^{2}}{\rho _{0}^{2}}%
\Theta \left( \rho _{0}-\rho \right) +\left( \rho -\rho _{0}\right) \right] .
\label{9}
\end{equation}

The metric given by Eq. (\ref{7}) is characterized by two parameters,
namely, the linear mass density $\mu $ and the linear angular momentum
density $j=\mu w\rho _{0}$. This approximate solution is justified in a
domain in which the Newtonian potential generated by the thin massive
cylindrical shell is much less than the unity which means that $\left|
a\left( \rho \right) \right| \ll 1$. The term with $b\left( \rho \right) $,
in this metric, being proportional to $j$ is completely due to the rotation
of the shell.

It is interesting to call attention to the fact that, in the weak field
approximation,  the Riemann curvature tensor outside the rotating shell is
completely determined by the function $a\left( \rho \right) $ only. The
contribution of the term with $b\left( \rho \right) $ is concentrated on the
shell itself. This means that, in the weak field approximation, the local
effects of curvature connected with the rotation of the shell are absent
outside it.

In order to solve the Dirac equation for a massless particle, given by Eq. (%
\ref{1}), in the spacetime of a rotating massive body given by the line
element (\ref{7}), we will choose the following set of tetrads
\begin{eqnarray}
e_{\left( 0\right) }^{\mu } &=&\left( 1+\frac{a}{4}\right) \delta _{0}^{\mu }%
\text{ };\text{ \ \ \ \ \ \ }e_{\left( 1\right) }^{\mu }=\left( 1-\frac{a}{2}%
\right) \delta _{1}^{\mu };  \label{10} \\
e_{\left( 2\right) }^{\mu } &=&-\frac{b}{\rho }\delta _{0}^{\mu }+\frac{1}{%
\rho }\left( 1-\frac{a}{2}\right) \delta _{2}^{\mu }\text{ };\text{ \ \ \ \
\ \ }e_{\left( 3\right) }^{\mu }=\left( 1-\frac{a}{2}\right) \delta
_{3}^{\mu }.  \nonumber
\end{eqnarray}
Computing the expression for $A_{\left( a\right) }$ and $B_{\left( a\right) }
$ we get
\begin{equation}
A_{\left( a\right) }=\left[ -\frac{1}{4}\frac{da\left( \rho \right) }{d\rho }%
+\frac{1}{2\rho }\left( 1-\frac{a\left( \rho \right) }{2}\right) \right]
\delta _{\left( a\right) }^{1}  \label{11}
\end{equation}
and
\begin{equation}
B_{\left( a\right) }=\frac{1}{4}\frac{b}{\rho ^{2}}\delta _{\left( a\right)
}^{3}  \label{12a}
\end{equation}
As we are considering only the gravitational field in the weak
approximation, it is not necessary to distinguish between global and
tetradic incides, so that the massless Dirac equation can be written as
\begin{equation}
i\gamma ^{\mu }\left( x\right) \left( \partial _{\mu }+A_{\mu }\left(
x\right) +i\gamma ^{\left( 5\right) }B_{\mu }\left( x\right) \right) \Psi
\left( x\right) =0.  \label{13a}
\end{equation}
Let us try a solution of Eq. (\ref{13a}) of the form
\begin{equation}
\Psi \left( x\right) =\left\{ \exp \left[ -\int_{c}^{x}dx^{\prime \mu
}\left( A_{\mu }\left( x^{\prime }\right) +i\gamma ^{\left( 5\right) }B_{\mu
}\left( x^{\prime }\right) \right) \right] \right\} \Psi _{0}\left( x\right)
\label{14}
\end{equation}
where $\int_{c}^{x}$ represents an integral along a path $c$ \ from a given
point $P$ up to the endpoint $x$. Thus, we get the following equation for $%
\Psi _{0}\left( x\right) $
\begin{equation}
\partial _{0}\Psi _{0}+\left( 1-\frac{3a}{4}\right) \gamma ^{\rho }\partial
_{\rho }\Psi _{0}+\frac{1}{\rho }\left[ \left( 1-\frac{3a}{4}\right) -b%
\right] \gamma ^{\varphi }\partial _{\varphi }\Psi _{0}+\left( 1-\frac{3a}{4}%
\right) \gamma ^{z}\partial _{z}\Psi _{0}=0,  \label{15}
\end{equation}
where $\gamma ^{\rho }$ and $\gamma ^{\varphi }$ are given by
\begin{eqnarray}
\gamma ^{\rho } &=&\cos \varphi \gamma ^{1}+\sin \varphi \gamma ^{2},
\nonumber \\
\gamma ^{\varphi } &=&-\sin \varphi \gamma ^{1}+\cos \varphi \gamma ^{2}.
\label{16}
\end{eqnarray}
Now, let us assume that the condition
\begin{equation}
\left( 1+\gamma ^{5}\right) \Psi _{0}=0  \label{17}
\end{equation}
is fulfilled. Equation (\ref{17}) together with Eq. (\ref{1}) constitute the
Weyl equations for a massless spin 1/2 particle. Thus, our four-dimensional
problem reduces to a bidimensional one. Thus, choosing the solution of \ Eq.
($\ref{15}$) as
\begin{equation}
\Psi _{0}=\left\{ \exp \left[ \left( -iEt+ikz+i\left( n+\frac{1}{2}\right)
\varphi \right) \right] \right\} \left(
\begin{array}{c}
u_{1}\left( \rho \right)  \\
u_{2}\left( \rho \right)
\end{array}
\right) ,  \label{18}
\end{equation}
we obtain the following equations for $u_{1}\left( \rho \right) $ and $%
u_{2}\left( \rho \right) :$
\begin{eqnarray}
&&\frac{d^{2}u_{1}\left( \rho \right) }{d\rho ^{2}}+\frac{1}{\rho ^{2}}\left[
\frac{3ka}{4\left( E-k\right) }\left( n+\frac{1}{2}\right) ^{2}-\frac{3ka}{%
4\left( E+k\right) }\left( n+\frac{1}{2}\right) \right.   \nonumber \\
&&\left. -\left( n+\frac{1}{2}\right) ^{2}\left( 1-a-2b\right) +\left( n+%
\frac{1}{2}\right) \left( 1-\frac{5a}{4}-b\right) -\frac{3a}{2}\left( n^{2}-%
\frac{1}{4}\right) \right.   \nonumber \\
&&\left. +\left( k-E\right) \left( 1+\frac{3ka}{4\left( E-k\right) }\right) +%
\frac{3a}{2}\left( \frac{k}{2}-E\right) \right] u_{1}\left( \rho \right)
\left. =0\right. ,  \label{19}
\end{eqnarray}
\begin{eqnarray}
&&\frac{d^{2}u_{2}\left( \rho \right) }{d\rho ^{2}}+\frac{1}{\rho ^{2}}\left[
\frac{3ka}{4\left( E-k\right) }\left( n+\frac{1}{2}\right) ^{2}+\frac{3ka}{%
4\left( E+k\right) }\left( n+\frac{1}{2}\right) \right.   \nonumber \\
&&\left. -\left( n+\frac{1}{2}\right) ^{2}\left( 1-a-2b\right) -\left( n+%
\frac{1}{2}\right) \left( 1-\frac{5a}{4}-b\right) +\frac{3a}{2}\left( n+%
\frac{1}{2}\right) \left( n+\frac{3}{2}\right) \right.   \nonumber \\
&&\left. +\left( k-E\right) \left( 1+\frac{3ka}{4\left( E-k\right) }\right) +%
\frac{3a}{2}\left( \frac{k}{2}-E\right) \right] u_{2}\left( \rho \right)
\left. =0\right. ,  \label{20}
\end{eqnarray}
whose  solutions are given by
\begin{equation}
u_{1}\left( \rho \right) =C_{1}\sqrt{\rho }J_{\frac{1}{2}\left| \nu \right|
}\left( \sqrt{\frac{D}{A}}\rho \right) +C_{2}\sqrt{\rho }N_{\frac{1}{2}%
\left| \nu \right| }\left( \sqrt{\frac{D}{A}}\rho \right)   \label{21}
\end{equation}
\begin{equation}
u_{2}\left( \rho \right) =D_{1}\sqrt{\rho }J_{\frac{1}{2}\left| \mu \right|
}\left( \sqrt{\frac{D}{A}}\rho \right) +D_{2}\sqrt{\rho }N_{\frac{1}{2}%
\left| \mu \right| }\left( \sqrt{\frac{D}{A}}\rho \right)   \label{22}
\end{equation}
where $C_{1}$, $C_{2}$, $D_{1}$ and $D_{2}$ are normalization constants, $%
\nu =\frac{\sqrt{\left( A+4B\right) A}}{A}$, $\mu =\frac{\sqrt{\left(
A+4C\right) A}}{A}$, with $A$, $B$ and $C$ given by
\begin{equation}
A=\left( E-k\right) ^{-1}\left[ \frac{3ka}{4\left( E-k\right) }-\left( 1-%
\frac{3a}{2}\right) \right] ,  \label{23}
\end{equation}

\bigskip
\begin{eqnarray}
B &=&\left( E-k\right) ^{-1}\left[ \left( n+\frac{1}{2}\right) ^{2}\left(
1-a-2b\right) \right.  \nonumber \\
&&\left. +\left( n+\frac{1}{2}\right) \left( 1-\frac{5a}{4}-b\right) -\frac{%
3ka}{4\left( E-k\right) }\left( n^{2}-\frac{1}{4}\right) \right] .
\label{25}
\end{eqnarray}

\begin{eqnarray}
C &=&\left( E-k\right) ^{-1}\left[ \left( n+\frac{1}{2}\right) ^{2}\left(
1-a-2b\right) \right.  \nonumber \\
&&\left. +\left( n+\frac{1}{2}\right) \left( 1-\frac{5a}{4}-b\right) -\frac{%
3ka}{4\left( E-k\right) }\left( n+\frac{1}{2}\right) \left( n+\frac{3}{2}%
\right) \right] ,  \label{24}
\end{eqnarray}

The constant $D$ which appears in the argument of the Bessel functions is
given by
\begin{equation}
D=k\left( 1-\frac{3a}{4}\right) -E.  \label{26}
\end{equation}

From the exponential factor in (\ref{14}) we conclude that $\exp \left[
-\int_{c}^{x}dx^{\prime \mu }A_{\mu }\left( x^{\prime }\right) \right] $
gives a contribution independent of the parameter $b$. On the other hand the
factor $\exp \left[ -i\int_{c}^{x}\gamma ^{\left( 5\right) }B_{\mu }\left(
x^{\prime }\right) dx^{\prime \mu }\right] $ gives a contribution $\exp
\left( -i\beta \gamma ^{\left( 5\right) }\right) $ where $\beta
=\int_{c}^{x} $ $B_{\mu }\left( x^{\prime }\right) dx^{\prime \mu }$ and
depends on the angular momentum of the source. In this case if one consider
a circularly polarized wave, the effect of this exponential factor is to
shift the phase of this wave by an angle $\pm \beta .$ If rather, one
consider a plane polarized wave, then the effect of the exponential factor $%
\exp \left( -i\beta \gamma ^{5}\right) $ is to rotate the plane of
polarization through an angle $-2\beta $.

Finally, if we consider the difference in phase or polarization between two
beams emitted by a source which follow two different paths in such a way
that they form a closed loop around the source, then the relative phase
shift depends on the angular momentum of the source.

From the solutions given by Eqs. (\ref{21}) and (\ref{22}) we conclude that
there is also a dependence on the parameter $b$. In summary the solution of
the Weyl equations depends on the angular momentum of the cylinder through
the phase factor of Eq. (\ref{14}) and the solution of Eq. (\ref{15}).

It is worth to call attention to the fact that in the region of motion of
the massless spin-$\frac{1}{2}$ particle, the Riemann curvature tensor, in
the linear approximation, does not depend on the angular momentum of the
cylinder of matter, but the solution is influenced by this quantity. This
result put into evidence a physical effect, namely, a generalization of the
gravitational Aharonov-Bohm effect, which means that even in the case
in which the particle is constrained to move
 in a region where the Riemann curvature does
not depend on the angular momentum of the source,
it exhibits a gravitational effect associated with this quantity.

\end{document}